\documentclass[11pt,preprintnumbers,aps,amssymb,nofootinbib,amsmath,superscriptaddress,notitlepage,prd]
{revtex4-1}
\usepackage{epsfig,epsf}
\usepackage{bm} 
\usepackage{color} 
\definecolor{darkgreen}{RGB}{0,128,0}
\definecolor{darkred}{RGB}{139,0,0}
\usepackage{slashed} 
\usepackage{soul} 
\usepackage{hyperref}
\newcommand{\beq}{\begin{equation}}
\newcommand{\beql}[1]{\begin{equation}\label{#1}}
\newcommand{\eeq}{\end{equation}}
\def\bal#1\gal{\begin{align}#1\end{align}}
\newcommand{\ball}[1]{\bal\label{#1}}
\newcommand{\eq}[1]{(\ref{#1})}
\newcommand{\fig}[1]{Fig.~\ref{#1}}
\renewcommand{\sec}[1]{Sec.~\ref{#1}}
\renewcommand{\b}[1]{{\bm #1}} 
\newcommand{\unit}[1]{\hat {{\bm #1}}} 

\newcommand{\Tr}{\operatorname{Tr}}

\newcommand{\e}{\varepsilon}

\newcommand{\sgn}{\operatorname{sgn}}
\begin{document}


\title{Chiral Cherenkov and chiral transition radiation in anisotropic matter}

\author{Kirill Tuchin}

\affiliation{Department of Physics and Astronomy, Iowa State University, Ames, IA 50011, USA}

\date{\today}


\begin{abstract}

 A significant contribution to the electromagnetic radiation by a fast electric charge moving in anisotropic chiral matter arises from spontaneous  photon radiation due to the chiral anomaly. While such a process, also known as the ``vacuum Cherenkov radiation", is forbidden in the QED vacuum, it can occur in chiral matter, where it is more appropriate to call it the ``chiral Cherenkov radiation". Its contribution to the radiation spectrum is  of order $\alpha^2$ compared to $\alpha^3$ of the bremsstrahlung. I derive the frequency spectrum and the angular distribution of this radiation in the high energy limit. The quantum effects due to the hard photon emission and  the fermion mass are taken into account. The obtained spectra are analyzed in the case the quark-gluon plasma and a Weyl semimetal.

\end{abstract}

\maketitle

\section{Introduction}\label{sec:i}

Electromagnetic radiation by fast particles in chiral matter has a number of unique features that make it a useful tool to investigate the chiral anomaly. The precision of experiments, especially in condensed matter physics, requires a quantitative description of the radiation. It can be derived by applying the standard machinery of the quantum field theory to the effective low energy Maxwell-Chern-Simons Lagrangian \eq{b1}. The chiral anomaly is encoded in the Chern-Simons term that couples electrodynamics to the pseudoscalar field $\theta$ which reflects the material topological properties. In quark-gluon plasma $\theta$  describes the gluon topological number fluctuations and the associated sphaleron transitions. In Weyl semimetals its gradient is proportional to the splitting between the Weyl nodes. The  Chern-Simons term induces a number of novel phenomena such as the chiral magnetic effect and the anomalous Hall effect \cite{Klinkhamer:2004hg,Zyuzin:2012tv,Grushin:2012mt,Kharzeev:2007tn,Kharzeev:2013ffa}. 

The first calculation of the electromagnetic radiation due to the $\theta$-term was done in \cite{Lehnert:2004hq,Lehnert:2004be} who dubbed it the ``vacuum Cherenkov radiation" and proposed as a test of the physics beyond the Standard Model. Specifically, it was proposed to be a test of the Lorentz symmetry violation \cite{Mattingly:2005re,Kostelecky:2002ue,Jacobson:2005bg,Altschul:2006zz,Altschul:2007kr,Nascimento:2007rb}. A phenomenon closely related to the vacuum Cherenkov radiation is the chiral transition radiation which is emitted  by fast particles crossing the boundary between the chiral matter and vacuum  \cite{Tuchin:2018sqe,Huang:2018hgk}.

The main goal of this paper is to re-derive the photon spectrum using the Lagrangian \eq{b1} with the from of the $\theta$-field dictated by the applications in condensed matter and nuclear physics. To this end, the calculation method of  \cite{Tuchin:2018sqe} is employed which allows one to take account of the quantum corrections due to the hard photon emission and the fermion mass. The paper is structured as follows. In \sec{sec:b} the effective Lagrangian is introduced. The resulting Maxwell-Chern-Simons equations effectively describe the free electromagnetic field in an anisotropic matter. It is assumed that $c_A\theta$, where $c_A$ is the anomaly coefficient,  is of order unity and thus the Chern-Simons $\tilde F F $ term is of the same order of magnitude as the usual $FF$ term of the Maxwell theory. The field is then quantized in \sec{sec:a} and its properties, in particular the polarization, are discussed.  The radiation rate is computed in \sec{sec:c} in the leading order in the perturbation theory.  Since in practical applications fermions are  ultra-relativistic, I employ the high-energy (or ultra-relativistic) approximation in \sec{sec:d}, which allows me to derive simple expressions for the photon spectra. The main results are given by \eq{d19},\eq{d24},\eq{d29} and \eq{d36}. Figures \ref{fig:1}--\ref{fig:3} and \ref{fig:4}--\ref{fig:5}  represent an application of the obtained results to a sample Weyl semimetal and to the quark-gluon plasma respectively. The summary is presented in \sec{sec:s}.

\section{Maxwell-Chern-Simons effective theory}\label{sec:b} 

The $CP$-odd domains in the chiral matter can be  described by a pseudoscalar field $\theta$ whose interaction with the electromagnetic field  $F^{\mu\nu}$ is governed by the Lagrangian \footnote{Throughout the paper the natural rationalized units $\hbar=c=1$, $\alpha=e^2/4\pi \approx 1/137$ are used.   
} \cite{Wilczek:1987mv,Carroll:1989vb,Sikivie:1984yz,Kalaydzhyan:2012ut}
\ball{b1}
\mathcal{L}= -\frac{1}{4}F_{\mu\nu}^2-\frac{c_A}{4}\theta\tilde F_{\mu\nu}F^{\mu\nu}+\bar \psi (i\gamma^\mu D_\mu-m)\psi\,,
\gal
where   $\tilde F_{\mu\nu}= \frac{1}{2}\epsilon_{\mu\nu\lambda\rho} F^{\lambda\rho}$ is  the dual field tensor and 
$c_A=N_c\sum_f q_f^2 e^2/2\pi^2$ is the chiral anomaly coefficient. In many applications, such as the Weyl semimetals,  $\theta$ is time-independent and has constant spatial gradient $\b\nabla \theta = \b b/c_A$. The same model can describe the spatially inhomogeneous $CP$-odd domains in the quark-gluon plasma \cite{Tuchin:2016qww}.  We will use this model throughout the paper.  The field equations of electrodynamics in chiral electrically neutral and non-conducting matter read 
\bal
&\b \nabla\cdot \b B= 0\,,  \qquad \b \nabla\cdot \b E=-\b b\cdot \b B \,, \label{b4}\\
& \b \nabla \times \b E= -\partial_t \b B\,,\qquad  \b \nabla \times \b B= \partial_t \b E+\b b\times \b E \,.\label{b6}
\gal
The monochromatic Fourier components of the electric and magnetic fields satisfy the equations \cite{Qiu:2016hzd}
\bal
&\b \nabla\cdot \b B= 0\,, \qquad \b \nabla\cdot \b D=0\label{b8}\\
& \b \nabla \times \b E= i\omega \b B\,,\qquad  \b \nabla \times \b B=-i \omega \b D \,.\label{b9}
\gal
where the displacement field $D_i= \varepsilon_{ij}E_j$ is given by
\ball{b13}
\b D= \b E+\frac{i}{\omega}\b b\times \b E=0\,.
\gal
Eqs.~\eq{b8},\eq{b9} describe free electromagnetic field in an anisotropic matter with the dielectric tensor 
\ball{b15}
\varepsilon_{ij}= \delta_{ij}- i\epsilon_{ijk}b_k/\omega\,.
\gal

Generally, the dispersive matter supports the transverse and longitudinal electromagnetic waves. However, since the dielectric tensor \eq{b15} does not depend on $\b k$, there is no spatial dispersion and hence no longitudinal waves.  The frequencies of the transverse waves can be found using the Fresnel equation
\ball{b17}
\left|k_ik_j-k^2\delta_{ij}+\omega^2_{\b k \lambda}\varepsilon_{ij}(\omega_{\b k \lambda})\right|=0\,.
\gal
Substituting \eq{b15} into \eq{b17} one finds (using e.g.\ \eq{a16}) 
\ball{b18}
\omega_{\b k\lambda}^2= k^2+\frac{b^2}{2}-\lambda \sgn(\b b\cdot \b k) \sqrt{\frac{b^4}{4}+(\b b\cdot \b k)^2}\,,
\gal
where $\lambda=\pm 1$ is the right/left-handed photon polarizations \footnote{In the limit $k\ll b$, the dispersion relation \eq{b18} has a gapped  $\omega_{\b  k -}=b$ and a gapless  $\omega_{\b  k +}=k\sin\beta$ branches, where $\cos\beta = \unit b\cdot\unit k$ (assumed to be positive). Only when $\beta=0$, one recovers the ``non-relativistic photon"  $ \omega_{\b  k +}= k^2/b$ reported in \cite{Yamamoto:2015maz}.  }. The corresponding group velocity is
\ball{b19}
v_{\b k\lambda} =  \frac{\partial\omega_{\b k\lambda}}{\partial k}= \frac{k}{\omega_{\b k\lambda}}\left( 1+ \frac{(\unit k \cdot \b b)^2}{\omega_{\b k \lambda}^2-k^2-b^2/2}\right)\,.
\gal
It is clearly different for the two photon polarizations. 


\section{Photon wave function in anisotropic dispersive medium}\label{sec:a} 

The electromagnetic field in anisotropic dispersive medium was quantized in \cite{AN}.  The corresponding expression 
in the radiation gauge is
\ball{a3}
\b A(\b x, t)=&\sum_{\b k\lambda}(a_{\b k \lambda}\b A_{\b k\lambda}+a^\dagger_{\b k \lambda}\b A_{\b k\lambda}^*)+\sum_{\b k\nu}( a_{\b k \nu} \b A_{\b k\lambda}+ a^\dagger_{\b k \nu}\b A_{\b k\lambda}^*)\,.
\gal
where $\lambda$ runs over the transverse polarizations whereas $\nu$ over the longitudinal ones. The corresponding wave functions read
\bal
\b A_{\b k\lambda}= &\b e_{\b k\lambda} \left( \frac{k\, v_{\b k\lambda}}{2\omega^2_{\b k\lambda}\varepsilon_{ij}e_{\b k\lambda i}^*e_{\b k\lambda j}V}\right)^{1/2} e^{i\b k \cdot \b x-i\omega_{\b k \lambda}t}\,,\label{a5}\\
\b A_{\b k\nu}=&\unit k \left( \frac{k^2}{\omega^2_{\b k\nu}k_ik_j \partial\varepsilon_{ij}/\partial \omega_{\b k\nu}V}\right)^{1/2}e^{i\b k \cdot \b x-i\omega_{\b k \nu}t}\,.\label{a6}
\gal
The creation and annihilation operators in \eq{a3} satisfy the usual bosonic commutation relations
\ball{a8}
[a_{\b k \lambda},a_{\b k' \lambda'}^\dagger]= \delta_{\b k \b k'}\delta_{\lambda \lambda'}\,,\quad
[a_{\b k \nu},a_{\b k' \nu'}^\dagger]= \delta_{\b k \b k'}\delta_{\nu \nu'}\,,\quad 
[a_{\b k \lambda},a_{\b k' \nu}^\dagger]= 0\,.
\gal
As indicated in \sec{sec:b}, a matter with the dielectric tensor \eq{b15} does not support the longitudinal waves. Therefore the second term in \eq{a3}, which is the vector potential operator of the longitudinal waves, vanishes.  

As for the transverse waves, for a given photon momentum  $\b k$, the transverse polarization vectors obey the system of equations
\ball{a10}
\left[k_ik_j-k^2\delta_{ij}+\omega^2_{\b k \lambda}\varepsilon_{ij}(\omega_{\b k \lambda})\right]e_{\b k \lambda j}=0\,.
\gal
They has non-trivial solutions labeled by $\lambda$ only if the  Fresnel equation \eq{b18} is satisfied.
The transverse polarization vectors $\b e_{\b k \lambda }$ satisfy the following conditions
\bal
&\varepsilon_{ij}k_ie_{\b k \lambda j}=0\,,\label{a13}\\
&\varepsilon_{ij}k_ie_{\b k \lambda i}^*e_{\b k \lambda' j}=\varepsilon_{ij}k_ie_{\b k \lambda i}^*e_{\b k \lambda j}\delta_{\lambda\lambda'}\,,\label{a14}
\gal
instead of the usual $\b k\cdot \b e_{\b k \lambda}=0$ and $\b e_{\b k \lambda }\cdot \b e^*_{\b k \lambda' }=\delta_{\lambda\lambda'}$. Eq.~\eq{a13} indicates that the displacement field $\b D$ is orthogonal to the wave vector $\b k$, whereas generally $\b E\cdot \b k \neq 0$.

To find the explicit from of the polarization vectors, consider a Cartesian frame with $z$-axis pointing in the direction of $\b b$. The dielectric tensor  \eq{b15} takes form
\ball{a16}
\varepsilon= \left(\begin{array}{ccc}1 & -ib/\omega_{\b k\lambda} & 0 \\ib/\omega_{\b k\lambda} & 1 & 0 \\0 & 0 & 1\end{array}\right)\,.
\gal
Its eigenvalues are $1\pm b/\omega$ and $1$ with the corresponding eigenvectors --- the principal dielectric directions --- given by  $\unit x_1=(\unit x+i\unit y)/\sqrt{2}$, $\unit x_2=(\unit y+i\unit x)/\sqrt{2}$ and $\unit x_3=\unit z$. To diagonalize the dielectric tensor we transform it to the ``principal coordinate system" span by these eigenvectors. Transformation matrix $P$ is given by
\ball{a18} 
P= \left(\begin{array}{ccc}1/\sqrt{2} & i/\sqrt{2} & 0 \\i/\sqrt{2} & 1/\sqrt{2} & 0 \\0 & 0 & 1\end{array}\right)\,.
\gal
One can verify that $\tilde\varepsilon =P^{-1}\varepsilon P$ is diagonal. In the principal coordinate system  \eq{a10} reads \cite{Qiu:2016hzd}
\ball{a20}
\left[k_i^*k_j-k^2\delta_{ij}+\omega^2_{\b k \lambda}\tilde \varepsilon_{ij}(\omega_{\b k \lambda}, \b k)\right]\tilde e_{\b k \lambda j}=0\,,
\gal
where $i,j =1,2,3$ label the orthogonal directions and we defined $k_1= (k_x-ik_y)/\sqrt{2}$, $k_2= (k_y-ik_x)/\sqrt{2}$, $k_3=k_z$. In particular, $k^2= k_1^*k_1+k_2^*k_2+k_3^2$. The frequencies $\omega_{\b k\lambda}$ are of course still given by \eq{b18}. The corresponding set of polarization vectors  is
\bal\label{a22}
\tilde{\b e}_{\b k \lambda}=C_{\b k \lambda} \left(\begin{array}{c} 
\frac{k_1/k}{(k/\omega_{\b k\lambda})^2-1-b/\omega_{\b k\lambda}} \\
\frac{k_2/k}{(k/\omega_{\b k\lambda})^2-1+b/\omega_{\b k\lambda}} \\
\frac{k_3/k}{(k/\omega_{\b k\lambda})^2-1}
\end{array}\right)\,.
\gal
 The polarization vectors in \eq{a5} are normalized so that   $\tilde{\b e}_{\b k \lambda}^*\cdot\tilde{\b e}_{\b k \lambda}=1$. This fixes the normalization constant $C_{\b k \lambda}$. We won't need the explicit expression for $C_{\b k \lambda}$  as it cancels out in the final expression for the photon spectrum. The polarization vectors in the original coordinate system are $ e_{\b k \lambda}=P \tilde e_{\b k \lambda}$.

\section{Radiation rate}\label{sec:c} 

The scattering matrix element for the transverse photon radiation $f(p)\to f(p')+\gamma(k)$ by a fermion of mass $m$ and electric charge $Q$ is given by
\ball{c3}
S=& -ie Q\int \bar \psi_{\b p' s'} \slashed{A}^*_{\b k \lambda} \psi_{\b p s}  d^4x\,.
\gal
Substituting the transverse photon wave function \eq{a5} and the electron wave function
\ball{c7}
\psi_{\b p s}(\b x, t) = \frac{1}{\sqrt{2\e V}}u_{\b ps}  e^{i\b p\cdot \b x-i \e t}\,, 
\gal
where $\e= \sqrt{p^2+m^2}$ one obtains
\bal
S
=&-ie Q(2\pi)^4\delta(\omega+\e'-\e)\delta(\b k+\b p'-\b p) \frac{\bar u_{\b p' s'}\slashed{e}^*_{\b k \lambda} u_{\b p s}}{\sqrt{8\e\e' \omega V^3}}\left( \frac{k v_{\b k \lambda}}{\omega\varepsilon_{ij}e_{\b k\lambda i}^*e_{\b k\lambda j}}\right)^{1/2}\,,\label{c5}
\gal
where we employed a shorthand notation $\omega= \omega_{\b k\lambda}$ and $\e'= \sqrt{p'^2+m^2}$. The radiation probability can be computed as
\ball{c8}
dw= \frac{1}{2}\sum_{\lambda s s'}|S|^2\frac{V d^3p'}{(2\pi)^3}\frac{V d^3k}{(2\pi)^3}\,,
\gal
implying that the rate is
\ball{c9}
dW= \frac{1}{2(2\pi)^2}e^2Q^2\sum_\lambda\delta(\omega+\e'-\e)\delta(\b k+\b p'-\b p) \frac{1}{8\e\e' \omega}
 \frac{k v_{\b k \lambda}}{\omega\varepsilon_{ij}e_{\b k\lambda i}^*e_{\b k\lambda j}} \sum_{ss'}|\mathcal{M}_0|^2 d^3p'\, d^3k\,,
\gal
where 
\ball{c11}
 \sum_{ss'}|\mathcal{M}_0|^2&= \Tr\left[(\slashed p+m) \slashed{e}^*_{\b k \lambda}(\slashed p'+m)\slashed{e}_{\b k \lambda}\right]\nonumber\\
 &=
 4e^*_{\b k \lambda i}e_{\b k \lambda j}\left[ p_ip_j'+p_jp_i'+\delta_{ij}(\e\e'-\b p\cdot \b p'-m^2)\right]\,.
\gal

The spectrum of photons emitted in a solid angle $d\Omega$ is 
\ball{c13}
 \frac{dW}{d\Omega d\omega}= \frac{\alpha Q^2}{16\pi}\sum_\lambda\delta(\omega+\e'-\e)\frac{k^3}{\e\e'\omega^2\varepsilon_{ij}e_{\b k\lambda i}^*e_{\b k\lambda j}}\sum_{ss'}|\mathcal{M}_0|^2\,.
 \gal

 \section{High energy limit}\label{sec:d} 
 
In practical applications it is useful to consider the high energy limit of the photon spectrum. The asymptotic form of the polarization vectors is obtained by expanding  \eq{a22} in the limit $b\ll k_z$. One finds  in the principle coordinate system 
\bal\label{d3}
\tilde{\b e}_{\b k \lambda}=\sqrt{\frac{k^2-k_z^2}{2k^2}} \left(\begin{array}{c} 
\frac{k_1}{k_z-\lambda k} \\
\frac{k_2}{k_z+\lambda k} \\
1
\end{array}\right)\,.
\gal
One can verify that, up to the terms proportional to the inhomogeneity parameter $\b b$, $\tilde{\b e}_{\b k \lambda}\cdot \b k=0$ as it must be in a homogenous matter. Transforming back to the original coordinate system yields
\ball{d5}
{\b e}_{\b k \lambda}= P\tilde {\b e}_{\b k \lambda}= 
\sqrt{\frac{k^2-k_z^2}{2k^2}} \left(\begin{array}{c} 
\frac{k_xk_z-i\lambda k k_y}{k_z^2-k^2} \\
\frac{k_zk_y+i\lambda k k_x}{k_z^2-k^2} \\
1
\end{array}\right)\,.
\gal
Taking now, for example, $k\approx k_x\gg k_y,k_z$ one obtains ${\b e}_{\b k \pm}\approx  \b \epsilon_{\b k\pm }= (\unit z\mp i \unit y)/\sqrt{2}$, i.e.\ a pair of mutually orthonormal  circular polarization vectors. 
Using the relation  ${\b \epsilon}_{\b k \pm}^*= {\b \epsilon}_{\b k \mp}$ and the fact that the expression in the square brackets in \eq{c11} is symmetric in $i,j$, one obtains in the high energy limit
\ball{d7}
 e_{\b k \pm i}^* e_{\b k \pm j}\to  \frac{1}{2}\left(  \epsilon_{\b k \pm i}^* \epsilon_{\b k \pm j}+ \epsilon_{\b k \pm j}^* \epsilon_{\b k \pm i}\right)= 
 \frac{1}{2}\left(  \epsilon_{\b k \pm i}^* \epsilon_{\b k \pm j}+ \epsilon_{\b k \mp i}^* \epsilon_{\b k \mp j}\right)
 =\frac{1}{2}\left(\delta^{ij}-\frac{k_ik_j}{k^2}\right)\,.
\gal
Substitution into \eq{c11} yields
\ball{d9}
 \sum_{ss'}|\mathcal{M}_0|^2=4\left[\e\e'-m^2- \frac{(\b k\cdot \b p)(\b k\cdot \b p')}{k^2}\right]\,.
\gal
In view of \eq{d7} and since $\varepsilon_{ij}e_{\b k\lambda i}^*e_{\b k\lambda j}\approx 1$, the dependence of the photon radiation rate \eq{c13} on $\b b$ arises entirely from the energy conserving delta-function. 

\begin{figure}[t]
      \includegraphics[height=4cm]{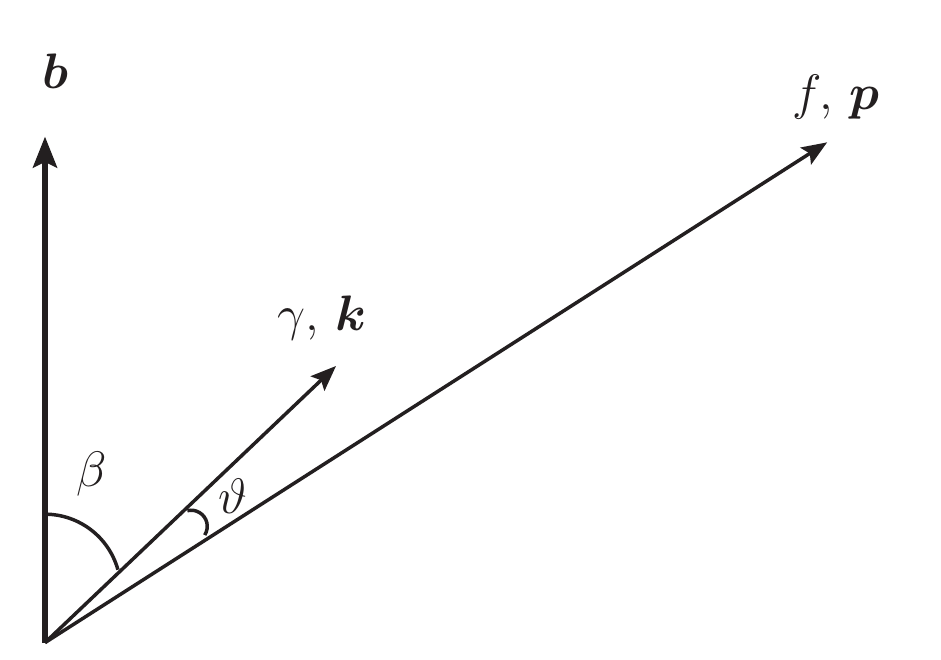}
  \caption{The geometry of the photon radiation. Since $\vartheta\approx k_\bot/\omega\ll 1$, the incident fermion and photon direction $\beta$ with respect to $\b b$ is almost the same.  }
\label{fig:G}
\end{figure}

Let $\b n$ be a unit vector in the direction of the incident momentum: $\b p = p\b n$.  In the high energy limit $k_\bot, \mu\ll k_n$ and $p'_\bot,m\ll p_n$, where $\mu$ is the ``photon mass" and  $\b k_\bot\cdot \b n= \b p'_\bot\cdot \b n=0$. Expanding 
\ball{d11}
k_z&\approx \omega \left( 1- \frac{k_\bot^2+\mu^2}{2\omega^2}\right)\,,\quad p_z'\approx \e'\left( 1- \frac{p'^2_\bot+m^2}{2\e'^2}\right)
\gal
and using the notation $x= \omega/\e$ one derives 
\ball{d13}
\sum_{ss'}|\mathcal{M}_0|^2=\frac{2}{x^2(1-x)}\left[k_\bot^2(2-2x+x^2)+m^2x^4\right]\,.
\gal
In the same approximation the delta-function can be written as 
\ball{d15}
\delta(\omega+\e'-\e)\approx 2x(1-x)\e \delta\left(k_\bot^2+\mu^2(1-x)+m^2x^2\right)\,,
\gal
where the photon effective mass is $\mu^2\approx -\lambda \omega b \cos\beta$ and the photon emission angle, see \fig{fig:G}
\ball{d17}
\cos\vartheta = \frac{\b k\cdot \b p}{kp}\approx 1-\frac{k_\bot ^2}{2xkp}\,,
\gal
so that $\vartheta \approx k_\bot/\omega$.  Using these equations in \eq{c13}  yields
\ball{d19}
\frac{dW}{d\Omega d\omega}= &\frac{\alpha Q^2x}{2\pi}\delta\left(x^2\e^2\vartheta^2+\kappa_\lambda\right)\left[\lambda\e b\cos\beta \left(1-x+\frac{x^2}{2}\right)-m^2 x\right]\theta\left(-\kappa_\lambda\right)\,,
\gal
where $\theta$ is the step function and 
\ball{d21}
\kappa_\lambda=\mu^2(1-x)+m^2x^2= -x(1-x)\lambda\e b \cos\beta+m^2x^2\,.
\gal
The parameter $\kappa_\lambda$ is negative if the following two conditions are satisfied:  $\lambda\cos\beta>0$ and 
$x<x_\text{max}$ where 
\ball{d22}
x_\text{max}= \left(1+\frac{m^2}{\lambda\e b \cos\beta}\right)^{-1}\,.
\gal
Thus, the photon polarization depends on the incidence angle, which is approximately the same as the angle $\beta$ between the photon direction and $\b b$: at $0<\beta<\pi/2$ it is right-handed, while at $-\pi/2<\beta<0$ it is left-handed. The spectrum of the right-handed photons  at a given incidence angle $\beta$ is the same as that of the left-handed ones at the incidence angle  $\pi-\beta$. 

\begin{figure}[ht]
      \includegraphics[height=5cm]{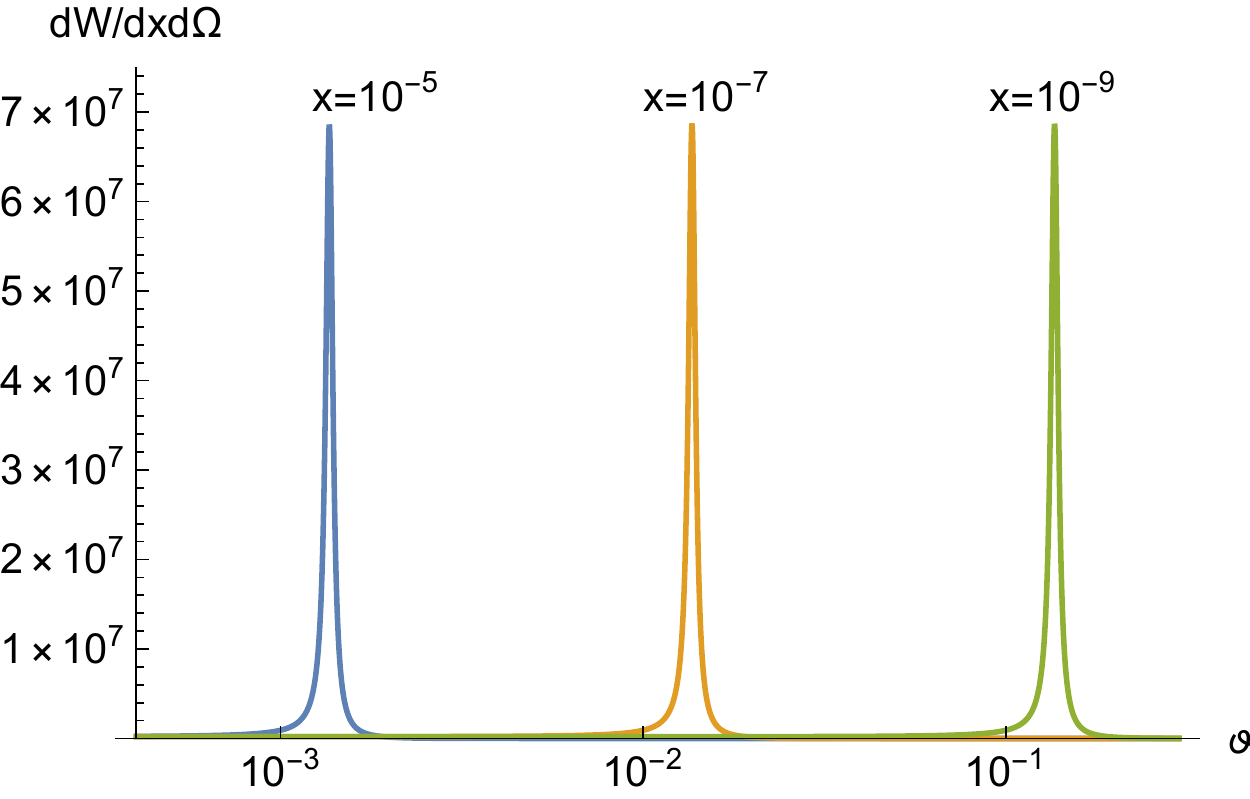}
  \caption{Angular distribution of  the right-hand photons with energies $\omega=x\e$ emitted by electron of energy $\e=10$~GeV moving through a chiral material with $b=0.19$~eV parallel to $\b b$ ($\beta=0$). The left-hand photons are not emitted. The delta-function in \eq{d19} is approximated by a Lorentzian of width $0.01\omega$. The maximum possible photon energy fraction is $x_\text{max}=7.4\cdot 10^{-3}$, see \eq{d22}.}
\label{fig:1}
\end{figure}

To illustrate the main features of the photon spectrum, I will choose a semimetal reported in \cite{Xu:2015cga,Lv:2015pya} that has $b= (\alpha/\pi) 80$~eV.  The photon radiation rate in a unit solid angle per unit frequency interval \eq{d19} is exhibited in \fig{fig:1} as a function of the emission angle. The sharp peaks correspond to the emission angles of photon with different frequencies and are located at 
\ball{d23}
\vartheta_0= \frac{\sqrt{-\kappa_\lambda}}{\omega}\approx \sqrt{\frac{\lambda b \cos\beta}{\omega}}\,,
 \gal
for small enough $x$.  

\begin{figure}[ht]
      \includegraphics[height=5cm]{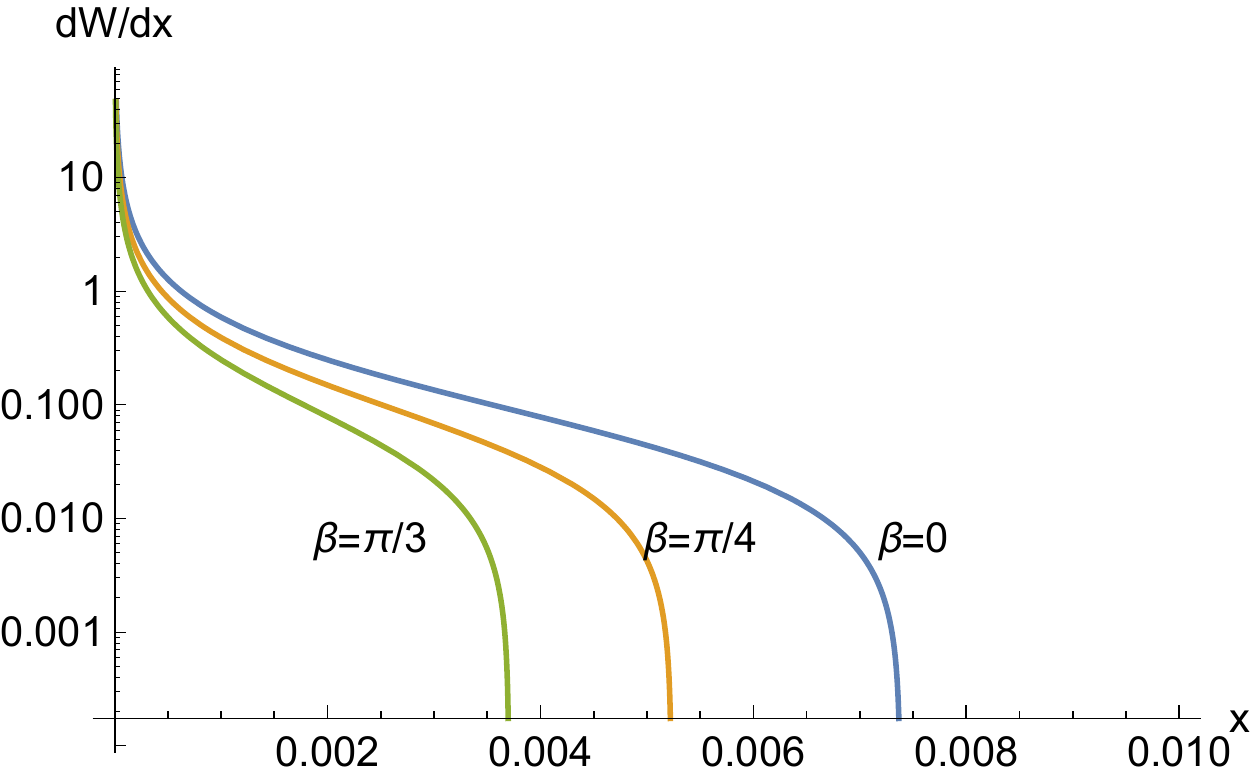}
  \caption{Spectra of the right-handed photons radiated by electron of energy 10~GeV moving through a chiral material with $b=0.19$~eV at different angles to $\b b$. The same exactly spectra for left-handed photons are obtained at angles $\beta=2\pi/3, 3\pi/4, \pi$ (left to right). }
\label{fig:2}
\end{figure}

Integration over the solid angle $d\Omega= \pi d\vartheta^2$ produces the energy spectrum
\ball{d24}
\frac{dW}{dx}= \frac{\alpha Q^2}{4\e} \frac{1}{x^2(1-x)}\left[ -(\mu^2(1-x)+m^2x^2)(2-2x+x^2)+m^2x^4\right]\theta\left(-\kappa_\lambda\right)\nonumber\\
= \frac{\alpha Q^2}{2\e x} \left[ \lambda \e b \cos\beta \left(1-x+\frac{x^2}{2}\right)-m^2x\right]\theta\left(-\kappa_\lambda\right)
\gal
in agreement with the earlier results \cite{Tuchin:2018sqe,Huang:2018hgk}. It is plotted in \fig{fig:2} for different incident angles $\beta$. The spectrum is dominated by soft photons $x\ll 1$ and terminates at $x=x_\text{max}$, which varies with $\beta$. 

\begin{figure}[ht]
      \includegraphics[height=5cm]{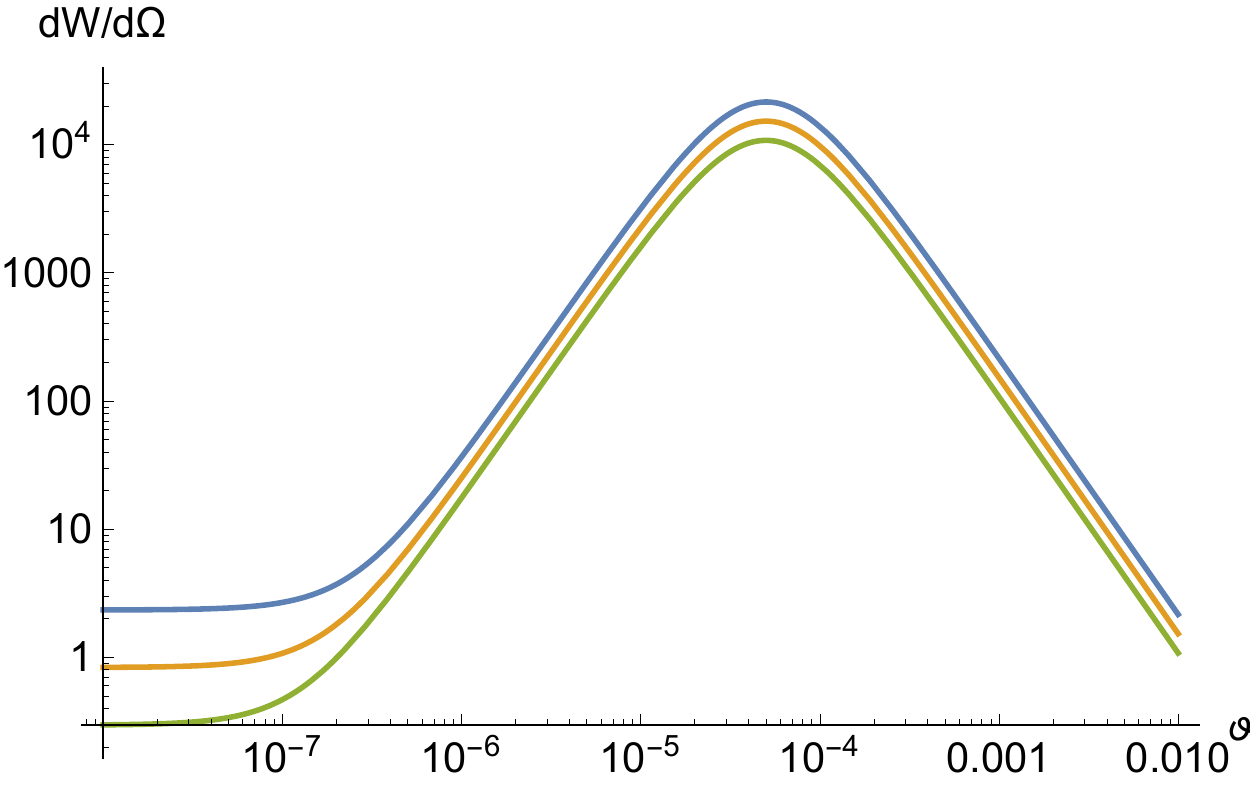}
  \caption{Angular distribution of the photon radiation produced by electron of energy 10~GeV moving through a chiral material with $b=0.19$~eV at the same angles $\beta$ to $\b b$ as in \fig{fig:2}. The maximum is at $\vartheta\approx m/\e$. }
\label{fig:3}
\end{figure}

The rate of photon emission into a solid angle is obtained by integrating \eq{d19} over the photon energies\footnote{Assuming that \eq{b18} holds throughout the entire photon spectrum. }
\ball{d29}
\frac{dW}{d\Omega}= \frac{\alpha Q^2x_0}{2\pi\lambda b \cos\beta}\left[\lambda\e b\cos\beta \left(1-x_0+\frac{x_0^2}{2}\right)-m^2 x_0\right]\theta\left(-\kappa_\lambda\right)\,,
\gal
where the emitted photon energy is fixed at 
\ball{d30}
x_0= \left(1+\frac{m^2+\e^2\vartheta^2}{\lambda\e b \cos\beta}\right)^{-1}\,.
\gal
One can verify that the rate \eq{d29} is proportional to $(m^2+\e^2\vartheta^2+b\e\lambda  \cos\beta)^{-3}$ implying that most of radiation is emitted into a cone with the opening angle 
\ball{d32}
\vartheta < \sqrt{\frac{b\e\lambda\cos\beta+m^2}{\e^2}}\,.
\gal
Since for the parameters of the benchmark model $b\e\ll m^2$, the radiation is emitted into a cone with the opening angle $\vartheta< m/\e$. This is shown in \fig{fig:3}. 

\begin{figure}[ht]
\begin{tabular}{cc}
      \includegraphics[height=5cm]{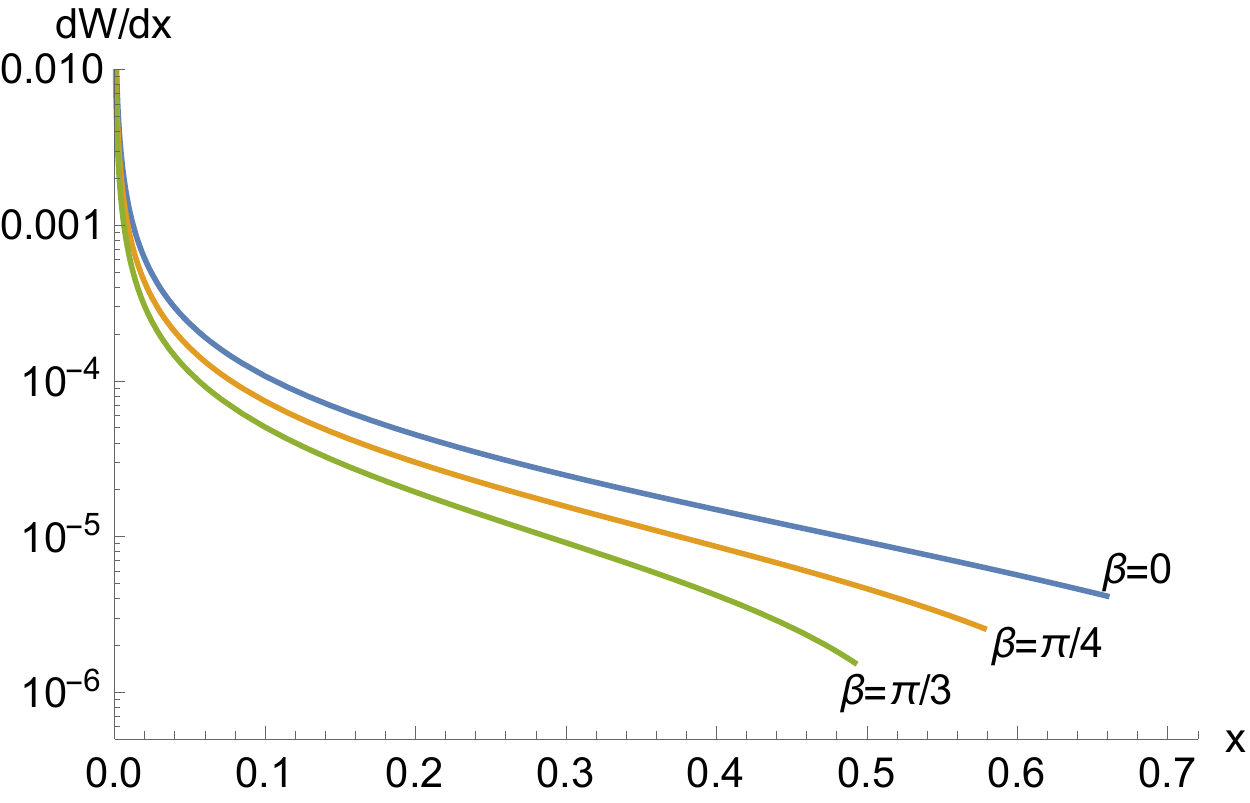} &
      \includegraphics[height=5cm]{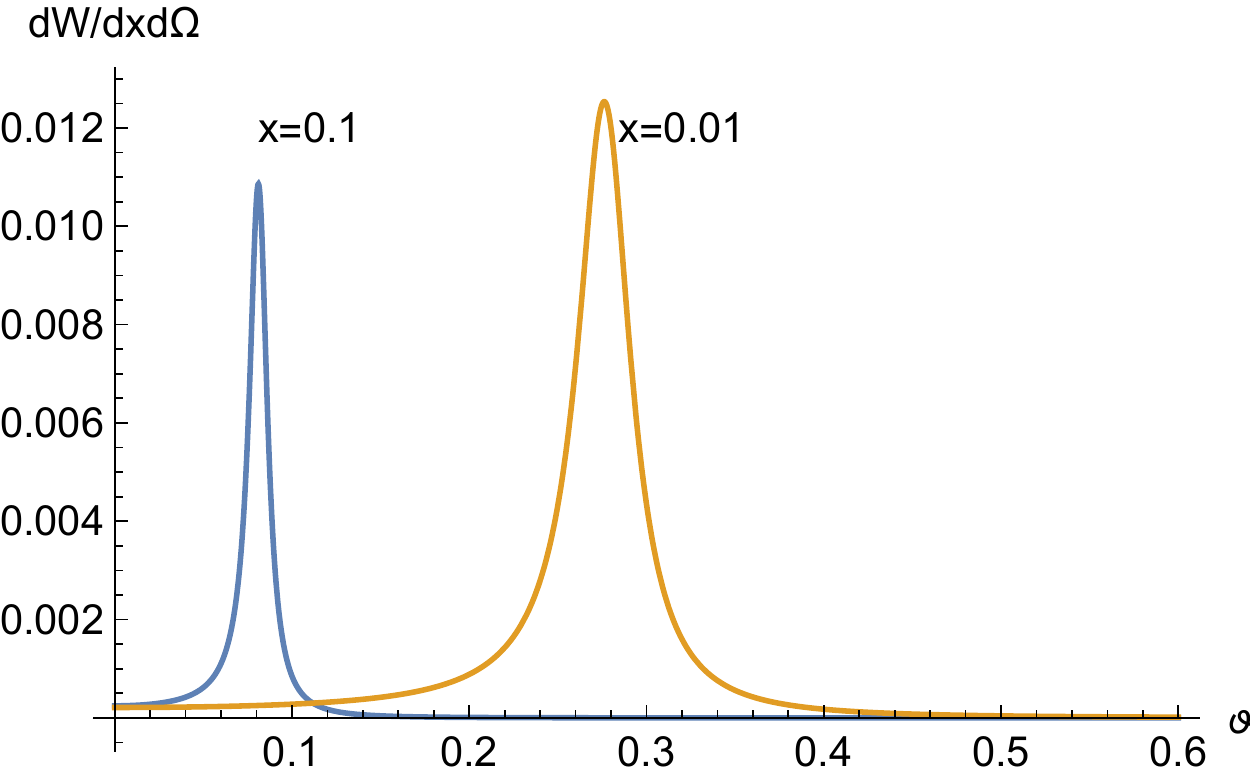}\\
      $(a)$ & $(b)$ 
      \end{tabular}
  \caption{Photon spectrum produced by a $u$-quark of energy 10~GeV and thermal mass 0.2~GeV moving in quark-gluon plasma. (a) Integrated over the photon emission angle for different incident angles $\beta$. (b) Angular distribution of the radiation for a fixed photon energy. In both plots the radiation is right-hand polarized.   }
\label{fig:4}
\end{figure}

As a different example, consider electromagnetic radiation by an ultra-relativistic $u$-quark in the quark-gluon plasma. The spatial gradients of $\theta$ can be reasonably estimated to be of order $\pi/R\sim 1$~GeV, where $R$ is domain linear size.  The corresponding spectra are shown in \fig{fig:4}.

The total photon radiation rate is derived by integrating \eq{d24} over $x$. The upper limit of integration is $x_\text{max}$ that ensures that $\kappa_\lambda<0$. The lower limit $x_\text{min}$ can be determined by recalling that the high energy approximation employed in this section assumes that $\omega\gg |\mu|$. This implies that  
\ball{d34}
x_\text{min}= \frac{b|\lambda  \cos\beta|}{\e}\,.
\gal
One therefore obtains in the logarithmic approximation (i.e.\ in the limit $x\ll 1$)
\ball{d36}
W= \int_{x_\text{min}}^{x_\text{max}}dx\frac{dW}{dx} = \frac{1}{2}\alpha Q^2b| \lambda  \cos\beta| \ln\left( \frac{m^2}{\e^2}+\frac{b|\lambda  \cos\beta|}{\e}\right)^{-1}\,.
\gal
The rate is only logarithmically dependent on the fermion energy $\e$.

\section{Transition radiation}\label{sec:e}

\begin{figure}[ht]
      \includegraphics[height=5cm]{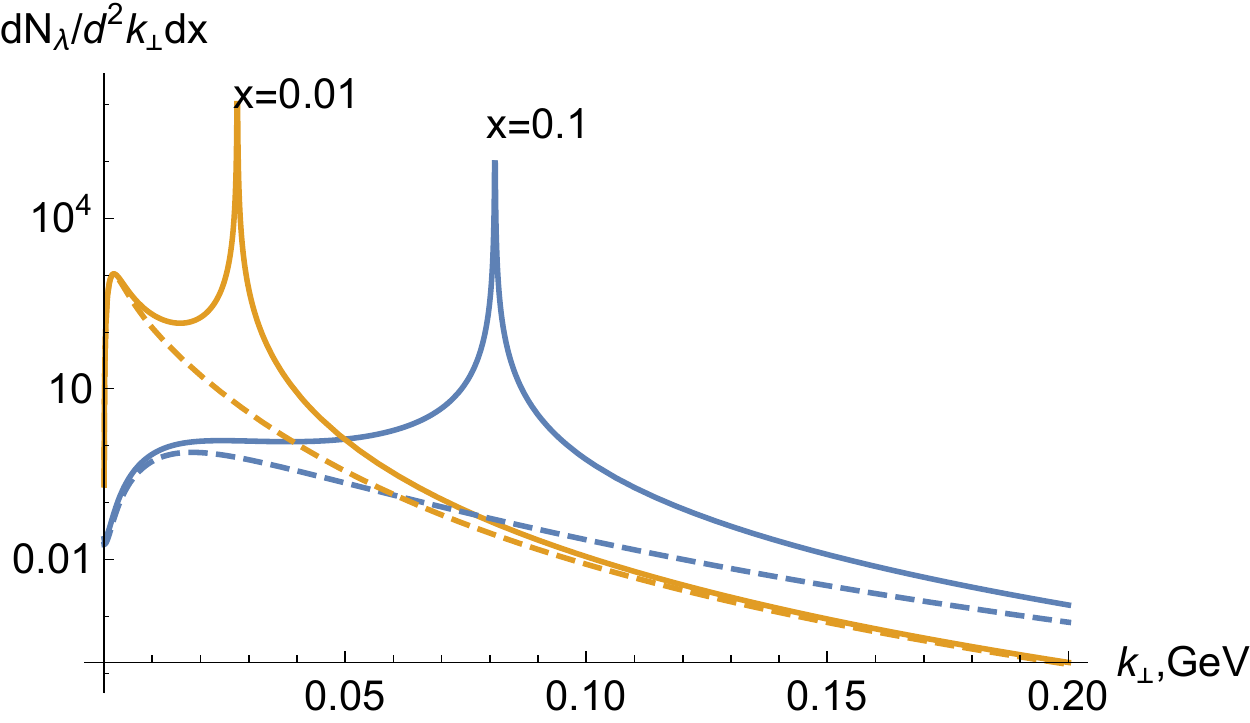}
  \caption{Photon spectrum emitted by a $u$-quark of energy 10~GeV and thermal mass 0.2~GeV moving in quark-gluon plasma at $\beta=0$. Solid lines: right-handed photons, dashed lines: left-handed photons. }
\label{fig:5}
\end{figure}

When a fermion radiating the electromagnetic radiation crosses the boundary between the chiral matter and  vacuum, it emits the chiral transition radiation discussed recently in \cite{Tuchin:2018sqe,Huang:2018hgk}. Assuming that the fermion is incident at the normal angle to the boundary and that there are no surface currents on it,  the spectrum of the radiated photons of a given polarization $\lambda$ is given by
\ball{e1}
\frac{dN_\lambda}{d^2k_\bot dx}= \frac{\alpha Q^2}{2\pi^2 x}\left\{ \left(\frac{x^2}{2}-x+1\right)k_\bot^2+\frac{x^4m^2}{2}\right\}
 \left[ \frac{1}{k_\bot^2+\kappa_\lambda}-\frac{1}{k_\bot^2+m^2x^2}\right]^2\,,
\gal
which coincides with the spectrum of the ordinary transition radiation when $\kappa_\lambda>0$ \cite{Baier:1998ej,Schildknecht:2005sc}.  The spectrum \eq{e1} has a resonance at $k_\bot  =-\kappa_\lambda$ corresponding to the delta-function in \eq{d19}. Thus, the formulas  for the spectra of the photon radiation derived in the previous section describe  this resonant behavior. In \fig{fig:5} one can see both the Cherenkov radiation peaks, which are the same as the peaks seen on \fig{fig:4}(b), and the transition radiation continuum. Unlike the chiral Cherenkov radiation, the chiral transition radiation is a boundary effect. 

 \section{Summary}\label{sec:s} 

The most important features of the photon spectrum observed in Figs.~\ref{fig:1}--\ref{fig:5} are: 
\begin{enumerate}

\item The angular distribution $dW/d\Omega d\omega $ peaks at certain emission angles $\vartheta_0$ that depends on photon energy $\omega$ and direction $\beta$ with respect to $\b b$, in agreement with the earlier observations \cite{Lehnert:2004hq,Lehnert:2004be}. 

\item The spectrum is always circularly polarized; the  polarization direction (right or left) depends on whether the angle $\beta$ is acute of obtuse. 

\item The spectrum $dW/dx$ terminates when the photon carries the maximum allowed fraction  $x_\text{max}$, given by \eq{d22}, of the fermion energy.

\item The total rate is proportional to $b$, but only logarithmically depends on the fermion energy. 

\end{enumerate}
I believe that these observations can make possible the experimental measurements of the novel electromagnetic radiation process in chiral materials. In particular, it can prove the existence of the $CP$-odd domains in quark-gluon plasma.

\acknowledgments
I am grateful to  Igor Shovkovy and Xu-Guang Huang for informative discussions. 
This work was supported in part by the U.S. Department of Energy under Grant No.\ DE-FG02-87ER40371.


\end{document}